%% file: main.tex
\pdfoutput=1
\documentclass[conference]{IEEEtran}
\IEEEoverridecommandlockouts
\usepackage{cite}
\usepackage{amsmath,amssymb,amsfonts}
\usepackage{algorithmic}
\usepackage{graphicx}
\usepackage{textcomp}
\usepackage{xcolor}

\usepackage[font=footnotesize]{caption}
\usepackage{subcaption}

\usepackage{makecell}
\usepackage{url}
\usepackage{glossaries}
\usepackage[utf8]{inputenc}
\usepackage{multicol}

\usepackage{booktabs}

\usepackage{soul}

\usepackage{tikz}
\usepackage{pgfplots}
\pgfplotsset{compat=1.16}
\usepgfplotslibrary{polar}
\usepackage{pgfplotstable}
\pgfplotsset{plot coordinates/math parser=false} 
\usetikzlibrary{plotmarks,patterns,patterns.meta,decorations.pathreplacing,backgrounds,calc,arrows,arrows.meta,spy,matrix}
\newlength\fheight
\newlength\fwidth

\DeclareUnicodeCharacter{2212}{−}

\makeglossaries
\loadglsentries{glossary}
\usepackage{hyperref}

\renewcommand{\figurename}{Fig.}
\usepackage[font=scriptsize]{subcaption}
\usepackage[font=footnotesize]{caption}

\newcommand{\rulesep}{\unskip\ \vrule\ }
\def\BibTeX{{\rm B\kern-.05em{\sc i\kern-.025em b}\kern-.08em
    T\kern-.1667em\lower.7ex\hbox{E}\kern-.125emX}}
    
\usepackage{dblfloatfix}
    
\begin{document}

\title{Autonomous Driving From the Sky:
\\Design and End-to-End Performance Evaluation}

\author{\IEEEauthorblockN{Matteo Bordin$^\star$, Marco Giordani$^\dagger$, Michele Polese$^\star$, Tommaso Melodia$^\star$, Michele Zorzi$^\dagger$}\vspace{0.1cm}
\IEEEauthorblockA{
$^\star$Institute for the Wireless Internet of Things, Northeastern University, Boston, MA, USA.}
\IEEEauthorblockA{
$^\dagger$Department of Information Engineering, University of Padova, Italy.}}

\maketitle

\maketitle

\begin{abstract}
For autonomous vehicles to operate without human intervention, information sharing from local sensors plays a fundamental role. 
This can be challenging to handle with bandwidth-constrained communication systems, which calls for the adoption of new wireless technologies, like in the \gls{mmwave} bands, to solve capacity issues. Another approach is to exploit \glspl{uav}, able to provide human users and their cars with an aerial bird's-eye view of the scene otherwise unavailable, thus offering broader and more centralized observations. 
In this article we combine both aspects and design a novel framework in which \glspl{uav}, operating at \glspl{mmwave}, broadcast sensory information to the ground as a means to extend the (local) perception range of vehicles. 
To do so, we conduct a full-stack end-to-end simulation campaign with ns-3 considering real UAV data from the Stanford Drone Dataset, and study four scenarios representing different \gls{uav}-to-ground communication strategies. 
Our results focus on the trade-off between centralized data processing in the sky vs. distributed local processing on the ground, with considerations related to the throughput, latency and reliability of the communication process.
\end{abstract}

\begin{picture}(0,0)(0,-360)
\put(0,0){
\put(0,0){\qquad \qquad \quad This paper has been submitted to IEEE for publication. Copyright may change without notice.}}
\end{picture}


\glsresetall

\section{Introduction}
The scientific community is witnessing an increasing interest in research and experimentation on autonomous driving vehicles, powered by the several benefits they provide (from improved safety to more efficient traffic management) and the market potential they generate~\cite{clements2017economic}.


For future vehicles to be fully autonomous, they will be equipped with diverse and heterogeneous sensors, from optical cameras to Light Detection and Ranging (LiDAR) sensors, able to perceive the environment and identify road entities in the surroundings~\cite{rossi2021role}.
In this scenario, more robust scene understanding can be achieved if vehicles share sensory data with other vehicles, which however imposes strict demands in terms of data rates, that may be difficult to support with legacy bandwidth-constrained communication systems~\cite{giordani2018feasibility}. 
One way to solve this issue is to compress and process the data before transmission~\cite{nardo2022point}, as well as to operate at high frequencies, e.g., in the \gls{mmwave} bands, where the large spectrum available, in combination with \gls{mimo} technologies, can support ultra-high transmission rates~\cite{9214394}.

At the same time, \glspl{uav}, mainly known as drones, have rapidly became popular thanks to the ease of deployment, low maintenance and operating costs, and native support for ubiquitous broadband coverage. 
When equipped with sensors, UAVs can enable several services, from crowd monitoring \cite{8761851} to airspace surveillance and border patrol \cite{8214963}. 
Drones have been further studied as a solution to provide connectivity to ground users and first responders in emergency situations \cite{8253543}, e.g., when cellular infrastructures are unavailable or no longer operational~\cite{8107663}. 

In recent years, \glspl{uav} have been also considered to support autonomous driving applications, especially for vehicular edge computing~\cite{Traspadini2022UAVHAPAssistedVE} and traffic management~\cite{lu2019uav}. In fact, UAVs operating from the sky can guarantee a birds'-eye wide perception of the scene than it would be possible from vehicles' (local) sensor acquisitions, thus achieving more centralized and precise observations.
Despite these benefits, however, the limited battery power and computational capacity available at the UAVs raise the questions of where to process and how to disseminate sensory data on the ground, in view of latency constraints.
Today, UAV communication is typically enabled by legacy wireless technologies such as Long Term Evolution (LTE)~\cite{7470934} which, however, may not satisfy the boldest latency and throughput requirements of future vehicular networks. In this respect, several prior works have demonstrated the feasibility of operating UAVs at \glspl{mmwave}\cite{8764406}, and characterized the optimal beamforming and deployment options for aerial nodes~\cite{wang2022beamforming}. 

\begin{figure*}[h]
\centering
    \begin{subfigure}{0.24\textwidth}
    \centering
        \includegraphics[width=0.5\textwidth]{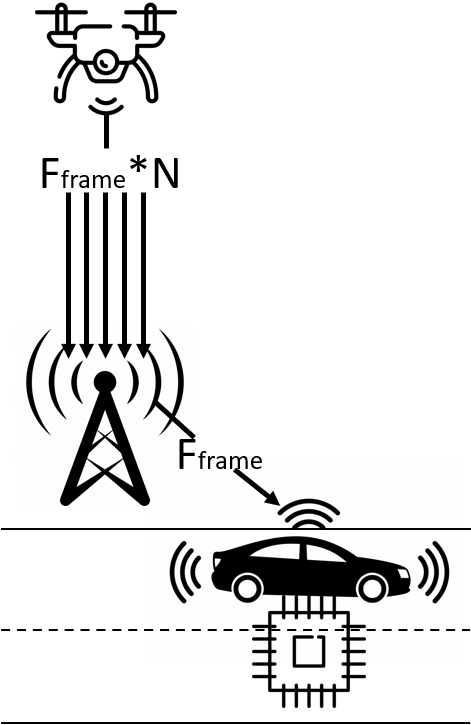} \caption{Multiple full frames (MFF)}
        \label{scenario1}
    \end{subfigure}
    \rulesep
    \begin{subfigure}{0.24\textwidth}
    \centering
        \includegraphics[width=0.5\textwidth]{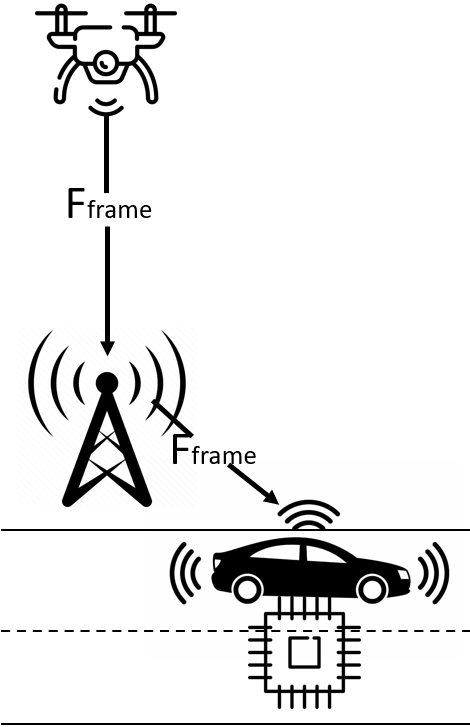} \caption{Broadcast full frames (BFF)}
        \label{scenario2}
    \end{subfigure}
    \rulesep
    \begin{subfigure}{0.24\textwidth}
    \centering
        \includegraphics[width=0.5\textwidth]{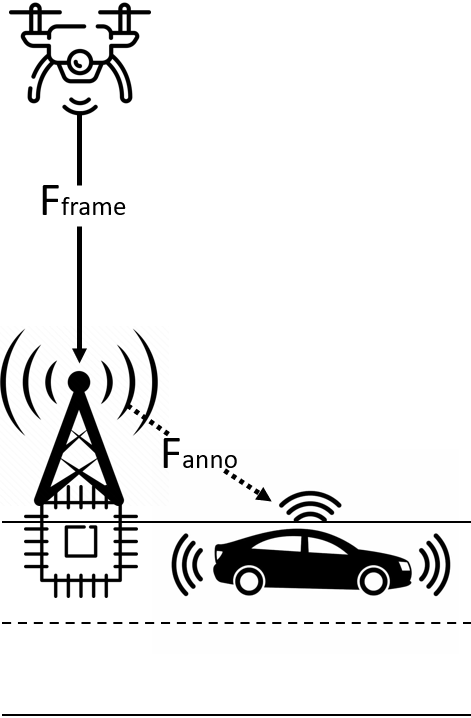} \caption{Broadcast frames and annotations (BFA)}
        \label{scenario3}
    \end{subfigure}
    \rulesep
    \begin{subfigure}{0.24\textwidth}
    \centering
        \includegraphics[width=0.5\textwidth]{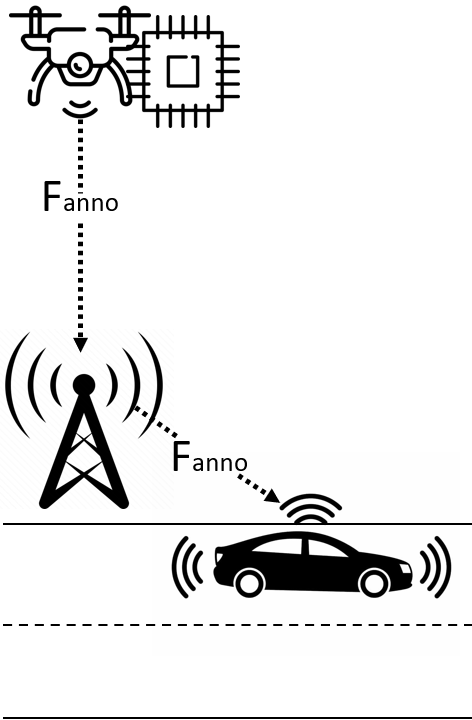} \caption{Broadcast annotations only (BAO)}
        \label{scenario4}
    \end{subfigure}
\caption{An illustration of the four UAV-to-ground communication scenarios. A chip icon is placed adjacent to the node that is performing the object detection.}
    \label{fig:scenarios}
\end{figure*}

Based on the above introduction, in this paper we evaluate the feasibility of implementing an autonomous driving framework by relying on \gls{uav}'s observations, and whether sensory information from the sky can be efficiently delivered to ground vehicles, possibly operating at \glspl{mmwave}.
To do so, we investigate several communication options between the \gls{uav} and the vehicles, each of which involves three main components: a \gls{uav} where sensory data are generated, a \gls{BS} acting as a relay, and multiple autonomous vehicles. Notably, we study whether autonomous driving tasks based on these data (e.g., object detection) should be processed on board of the UAV, or delegated to on-the-ground nodes.
The performance of the different schemes will be evaluated in ns-3 using the \texttt{mmwave} module~\cite{8344116} and real-world UAV data collected in the Stanford Drone Dataset~\cite{robicquet2016learning}, which promotes extreme levels of realism and permits to analyze the network considering full-stack end-to-end metrics and a real-world dataset as input. 
Our preliminary results demonstrate that data processing at the BS guarantees more efficient communication, in view of the limited power and computational capabilities of the UAV.

The rest of this paper is organized as follows. 
In Sec.~\ref{stateofart} we review the most recent works on UAV-based autonomous driving research, 
in Sec.~\ref{systemmodel} we present our system model and communication scenarios, in Sec.~\ref{scenimplementation} we describe how we extended \gls{ns-3} to simulate UAV-to-ground communication, in Sec.~\ref{results} we present our main numerical results, whereas conclusions are summarized in Sec.~\ref{conclusion}.

\section{State of the art} \label{stateofart}
In this section, we discuss the recent research in the area of UAV-based autonomous driving, specifically computational offloading and data dissemination. 
Hayat \emph{et al.}, in \cite{9363523}, evaluate the burden of data (image) processing for UAV autonomous navigation, that can be done on board, or fully/partially offloaded to an edge server. Similarly, in \cite{10.1145/3341568.3342109} the authors study the performance of UAV edge computing using Hydra, an architecture for the establishment of flexible sensing-analysis-control pipelines over autonomous airborne systems. 

If cellular infrastructures are unavailable (e.g., damaged by natural disasters), data offloading can be also between ground vehicles, whose limited computing and energy resources make it difficult to execute computationally sensitive mobile applications on board, and aerial platforms. In \cite{9214892}, the author suggest to offload computing tasks from the ground to UAVs that carry edge serves, and propose an algorithm to minimize the total energy and time required for the UAVs to complete the offloaded tasks, while optimizing their 3D flying height and horizontal positions.
Computational offloading may be also assisted by high altitude platforms (HAPs), as proposed in \cite{9621158}, where the authors designed a framework to offload communication and computational resources to aerial nodes to maximize the total number of user device requests with satisfied delay requirements while minimizing the total energy consumption.
Similarly, in our prior work \cite{Traspadini2022UAVHAPAssistedVE}, we formalized an optimization problem in which tasks are modeled as a Poisson arrival process, and applied queuing theory to identify how ground vehicles should offload resource-hungry tasks to UAVs, HAPs, or a combination of the two.

With respect to the state of the art, in this paper we do not focus on computational offloading, but rather on how UAV data are disseminated to ground vehicles, and where to process them. Moreover, while most literature focuses on UAV-to-ground communication in the legacy bands, and/or considers link-level evaluations, we perform end-to-end simulations in ns-3 considering mmWave frequencies, as well as both on-board and fully-offloaded computation.

\section{UAV-to-Ground Communication Scenarios} \label{systemmodel}

In this section we present four possible strategies for UAV-to-ground communication, as illustrated in Fig.~\ref{fig:scenarios}. Notably, each scenario consists of the following elements: a UAV that is recording videos from the sky, $N$ autonomous cars on the ground, and a \gls{BS} (or gNB, in 5G NR parlance) forwarding data from the UAV to the cars.
The UAV is placed at the center of the scene (e.g., at a road intersection) at height $h$, the BS is placed on the ground, perpendicular to the UAV in order to maintain a stable connection, and vehicles are allocated randomly within a rectangle.

 The four models differ in the way the data are broadcast, and the location of the computing platform (the chip icon in Fig.~\ref{fig:scenarios}) where UAV sensory data is processed to detect critical road entities in the scene.  

\paragraph{Scenario 1 --- \gls{MFF}} 
In the first scenario (Fig.~\ref{scenario1}) the drone is sending frames via the BS at a rate $F_{\rm frame}$ to each ground vehicle, that will eventually perform object detection using its own on-board computational capacity. 
The frame rate is not optimized so, in a situation where all packets are delivered without errors (best-case scenario), the total frame rate in the first link (UAV-\gls{BS}) would be equal to $NF_{\rm frame}$, i.e., the sum of the frames sent in each second link (BS-vehicle). 
On the downside, computation on board of vehicles may incur non-negligible delays given the limited capacity of budget vehicles, and the data rate in the first link would be $N$ times larger than the total data rate of each second link. In turn, this approach does not require coordination with the BS.

\paragraph{Scenario 2 --- \gls{BFF}} 
In the second scenario (Fig.~\ref{scenario2}), frames are sent at an optimized rate. 
While in the \GLS{MFF} scenario sensory data in the first link were replicated $N$ times, with optimized settings the UAV sends only one frame to the \gls{BS}, that will create $N$ copies of the received packets and eventually forward them to each of the $N$ cars. Finally, each vehicle will perform object detection on the received data, which may still incur long delays due to computational limitations. Ideally, the throughput in the first link would be equal to the throughput of each second link. In other words, the ideal throughput of the first link is $N$ times smaller than the sum of the throughput of all the second links.

\paragraph{Scenario 3 --- \gls{BFA}} 
In the third scenario (Fig.~\ref{scenario3}) the UAV sends one copy of all the frames to the \gls{BS}, which then performs object detection on the received data. 
This approach promotes faster processing than in the previous scenarios, as BSs are typically connected to continuous power sources and do not pose strict limitations in terms of computational capacity, space and storage. Eventually, the processed output (i.e., the bounding boxes of the detected objects, also referred to as \emph{annotations}) is returned to the ground vehicles at a frame rate $F_{\rm anno}$, in a packet of a much smaller size than the original frame,  which allows to reduce the communication latency on the second links.
In particular, the size of an annotation $\alpha$ is calculated as $\alpha=N\beta$, where $\beta$ is the memory size of a bounding box and $N$ is the number of detected objects.
To find the value of $\beta$, we made offline simulations to generate bounding boxes from real-world UAV video recordings collected in the Stanford Drone Dataset~\cite{robicquet2016learning}. To do so, we used the YoloV5 algorithm~\cite{11539}, a common benchmark in this field.

\paragraph{Scenario 4 --- \gls{BAO}} 
In the fourth scenario (Fig.~\ref{scenario4}), object detection is performed as soon as the frame is generated, i.e., on board of the UAV. While this allows low-size annotations to be sent already on the first link, as well as on the second link, thereby reducing the overall communication latency, the computational capacity of aerial nodes is generally lower than that available at the BSs, which may increase the processing delay compared to the \GLS{BFA} scenario. Ideally, the per-user throughput in the first link is equal to the per-user throughput in each second link. 

\section{ns-3 Implementation} \label{scenimplementation}
In this section we describe how we extended the ns-3 simulator to implement the four communication scenarios presented in Sec.~\ref{systemmodel}. 
While most simulators focus on \gls{phy} and \gls{mac} layer designs, and sacrifice the accuracy of the higher layers to reduce the computational complexity, ns-3 incorporates accurate models of the whole protocol stack, thus enabling scalable end-to-end simulations. 
In particular, in our work communication nodes operate at \glspl{mmwave}. As such, we use the \texttt{ns3-mmwave} module, described in~\cite{8344116}, which enables
the simulation of 5G-NR-compliant end-to-end cellular networks at \gls{mmwave} frequencies. 
It features a complete stack for \glspl{ue} and gNBs, with custom PHY (described in \cite{7805314}) and \gls{mac} layers with an \gls{OFDM} frame structure, dynamic \gls{TDD}, \gls{AMC}, and several scheduler implementations. Thanks to the integration with ns-3, it also features a complete implementation of the \gls{udp} stack. The simulator also features a 3GPP-compliant channel model, as well as antenna and beamforming models for mmWave communications~\cite{zugno2020implementation}. 
In terms of the implementation, in the \GLS{MFF} scenario \gls{udp} is installed at the end vehicles: the UAV is set up as a client while vehicles as servers. 
With this configuration, the UAV sends the same amount of packets $NF_{\rm frame}$ to every car. On the other hand, to implement broadcast communications in the other scenarios, some changes were applied to the \texttt{ns3-mmwave} module that manages the forwarding of the packets at the BS. With these changes, the UAV is sending only one copy of each packet to the BS, that in turn produces $N$ copies of the received packet, which are transmitted to the vehicles on the ground.

Finally, the packet size and the packet sending rate are set. For the simulations where the size of a UAV frame is larger than the maximum size of a UDP packet (UDP$_{\rm pck}$), the data must be split and sent in smaller packets of size UDP$_{\rm pck}$. 
In \GLS{MFF} and \GLS{BFF} and in the first link of \GLS{BFA}, the packets sending rate is equal to $1/((i \cdot F_{\rm frame})/\text{UDP}_{\rm pck})$ where $i$ is the total size in Byte of the sensory frame to be sent, and $F_{\rm frame}$ is the source frame rate. 
For the second link in \GLS{BFA}, and in \GLS{BAO}, the annotation rate is $1 / (F_{\rm anno} / (\beta \cdot N))$.

\section{Performance evaluation} \label{results}

In this section we introduce our performance evaluation setup, and discuss the simulation performance of the different UAV-to-ground communication scenarios.

\begin{table}[t]
\centering
\vspace{0.1in}
\renewcommand{\arraystretch}{1}
\caption{Main simulation parameters.}\label{table:parametersoverview}
\begin{tabular}{ll}
\toprule
Parameter & Value \\
\midrule
Carrier frequency & 28 GHz\\
Bandwidth & 1 GHz\\
BS TX power & 30 dBm\\
UAV TX power & 30 dBm\\
RLC buffer size & 10 MB \\
Frame rate & $\{15, 30\}$ FPS \\
Frame size & See Fig.~\ref{fig:SizeFrame} \\
Number of UEs & $\{4, 21\}$ \\
Simulation time & 15s \\
\bottomrule
\end{tabular}
\end{table}

\subsection{Simulation Parameters}

\setcounter{figure}{2}
\begin{figure*}[b]
\centering
\begin{subfigure}[t]{0.31\linewidth}
\centering
\setlength\fwidth{.8\columnwidth}
\setlength\fheight{.5\columnwidth}
    \input{images/ThroughputEndToEndUSerX.tex}
    \setlength\abovecaptionskip{-.3cm}
    \setlength\belowcaptionskip{-.1cm}
    \caption{Per-user throughput.}
    \label{fig:throughputComparison}
\end{subfigure}
\hfill
\begin{subfigure}[t]{0.31\linewidth}
\centering
\setlength\fwidth{.8\columnwidth}
\setlength\fheight{.5\columnwidth}
    \input{images/avg_latencyEndToEndUSerX.tex}
        \setlength\abovecaptionskip{-.3cm}
    \setlength\belowcaptionskip{-.1cm}
    \caption{Per-user latency.}
    \label{fig:latencyComparison}
\end{subfigure}
\hfill
\begin{subfigure}[t]{0.31\linewidth}
\centering
\setlength\fwidth{.8\columnwidth}
\setlength\fheight{.5\columnwidth}
    \input{images/Reliability_pckt_EndToEndUSerX.tex}
        \setlength\abovecaptionskip{-.3cm}
    \setlength\belowcaptionskip{-.1cm}
    \caption{Per-user reliability.}
    \label{fig:reliabilityComparison}
\end{subfigure}
\hfill
\caption{Performance evaluation for the 4 communication scenarios, with two different frame rates. MFF stands for multiple full frames, with the UAV sending one frame for each vehicle, then relayed by the BS. BFF stands for broadcast full frames, with the UAV sending a common reference frame for all vehicles, then relayed by the BS. With broadcast frames and annotations, or BFA, the UAV sends the common reference frame, and the BS forwards only the annotations. Finally, with broadcast annotations only, or BAO, the UAV sends annotations which are then relayed by the BS.}
\label{fig:comparison}
\end{figure*}

\setcounter{figure}{1}
\begin{figure}[t]
\centering
\setlength{\belowcaptionskip}{-0.5cm}
\setlength\fwidth{.75\columnwidth}
\setlength\fheight{.3\columnwidth}
\input{images/FrameSize.tex}
\caption{Size of each frame vs. the number of detected objects/vehicles.}
\label{fig:SizeFrame}
\end{figure}

Our simulator implements the communication scenarios described in Sec.~\ref{systemmodel}. The main parameters for the simulations are described in Table~\ref{table:parametersoverview}. We also provide the  source code and simulation scripts as a reference.\footnote{\url{https://bitbucket.org/mat_bord/autonomous-driving-from-the-sky}}
We run the simulations with the ns-3 Simulation Execution Manager (SEM) library~\cite{magrin2020sem}, which takes care of running multiple statistically independent instances of the same scenario and collecting relevant metrics. The ns-3 simulation time is set to 15 s, and we consider the following parameters: 
\begin{itemize}
    \item \emph{Application frame rate}. According to the Stanford Drone Dataset, the camera of the UAV records at $F_{\rm frame} \in \{15,30\}$ frames per second (FPS). Therefore, we set the frame rate of the ns-3 application generating UAV data to $1/F_{\rm frame}$ (independently on weather annotations or full frames are transmitted).
    \item \emph{Number of vehicles ($N$)}. It corresponds to the number of objects detected in each frame of the Stanford Drone Dataset videos, and sets the number of vehicles in each simulation. Based on offline simulations, we obtained that each processed frame featured from 4 to 21 vehicles, as shown in Fig.~\ref{fig:SizeFrame}.
    \item \emph{Full frame size}. It is the size of a frame of the Stanford Drone Dataset videos to be sent from the UAV, as shown in Fig.~\ref{fig:SizeFrame}. Notice that the frame size does not follow an increasing trend with the number of vehicles in the scene, i.e., the size of each full frame does not necessarily correlate with the number of vehicles. 
    \item \emph{Annotation size $\alpha$}. It is the size of an annotation produced after object detection. It is modeled as $\alpha=\beta N$, where $\beta = 39.7$ bytes is the average size of a single bounding box detected by the YoloV5 detection algorithm~\cite{11539}, and $N$ is the number of vehicles in the scene.
\end{itemize}

We consider 90 random UAV frames from the Stanford Drone Dataset, and run multiple statistically independent simulations to capture the following end-to-end metrics:

\begin{itemize}
    \item \emph{Per-user throughput.} It corresponds to the total number of received bytes per user divided by the total simulation time, averaged over all connected vehicles.
    \item \emph{Per-user reliability.} It is measured as the ratio between the number of packets delivered to the cars without errors and the total number of packets  transmitted by the~UAV.
    \item \emph{Per-user latency.} It is modeled as $L1 + L2$, where $L1$ represents the latency between the UAV and the BS (uplink) and $L2$ represents the latency between the BS and the vehicles (downlink), averaged over all connected vehicles. Both $L1$ and $L2$ account for the transmission time as well as the queuing time resulting from NR-specific scheduling and buffering, as modeled in the \texttt{ns3-mmwave} module.
\end{itemize}

\subsection{Performance Evaluation}

In the following paragraphs we compare the performance of the different communication scenarios described in Sec.~\ref{systemmodel}. 
Fig.~\ref{fig:comparison} reports the end-to-end throughput, latency and reliability for all configurations.

\textbf{\GLS{MFF} scenario.} 
The results in Fig.~\ref{fig:comparison}  clearly highlight that the wireless network (despite a bandwidth of 1 GHz) cannot support \gls{MFF} with more than 11 vehicles and 30 FPS. Notably, the throughput (Fig.~\ref{fig:throughputComparison}) and reliability (Fig.~\ref{fig:reliabilityComparison}) decrease, while the latency (Fig.~\ref{fig:latencyComparison}) increases to more than 200 ms. This is due to the a bottleneck in the first link (UAV-BS), which transmits $N$ times more data compared to each second link (BS-car). 
Notice that the latency of the second link is not particularly representative, as it is relative to only the correctly received packets. Given that most packet losses happen on the first link, which makes the system less congested, the (few) packets that make it to the second link are then transmitted with very low latency.
This also explains the latency plateau at around 200 ms, due to the fact that the UAV transmit buffer (e.g., at the \gls{rlc} layer) overflows for more than 11 cars.

On the other hand, the \gls{MFF} configuration can better support an application generating data at 15 FPS, as a consequence of the 50\% less traffic on the UAV-BS link, and the resulting less populated RLC queues at the UAV. 
The system performance is stable for up to 19 vehicles (Fig.~\ref{fig:throughputComparison}). After this threshold the UAV buffer saturates causing a degradation in latency (which reaches the 200 ms plateau) and reliability.

\textbf{\GLS{BFF} scenario.} The \gls{BFF} strategy is more efficient than \gls{MFF}, as it does not saturate the UAV buffer and the capacity of the first link by avoiding unnecessary duplication of the frames in the uplink. The performance of \gls{BFF} with an application rate of 30 FPS only degrades for more than 20 connected vehicles. 
Unlike MFF, this is due to a saturation of resources in the downlink, i.e., in the second links from the BS to the end vehicles. 
In fact, Fig.~\ref{fig:SizeFrame} shows that the file size of a frame with few users (e.g., 4) is comparable with the file size of a frame with 21 users. In the first case, however, the resources on the wireless link are split only among the UAV (uplink) and 4 more users (downlink), while in the latter more than 20 users contend for the same downlink resources, which may saturate the available capacity. BFF also easily sustains the performance with 15 FPS at the application.

This strategy, while being more efficient than MFF, requires the support for multicasting support at the RAN, a feature that has been only recently standardized in 5G networks~\cite{araniti2017multicasting,zeng2021field}.

\textbf{\GLS{BFA} scenario.}\label{scenario3results} 
%
%
BFA, which assumes data processing at the BS and the transmission of annotations in the downlink, manages to easily support the traffic for all the users in the tested environment. Fig.~\ref{fig:reliabilityComparison} and Fig.~\ref{fig:latencyComparison} show that this scheme provides ultra-high reliability, with an average of 98.472\% correctly delivered annotations, and an end-to-end latency as low as 3 ms, respectively. In this case, the performance with 30 and 15 FPS is comparable.
Notice that the BFA throughput is reasonably lower than that of MFF and BFF. This is due to the much lower size of annotations compared to full frames (on average up to 4 orders of magnitude), which limits the source rate on the second link. 

\setcounter{figure}{3}
\begin{figure}[t]
     \centering
     \setlength\fwidth{.75\columnwidth}
\setlength\fheight{.5\columnwidth}
     \input{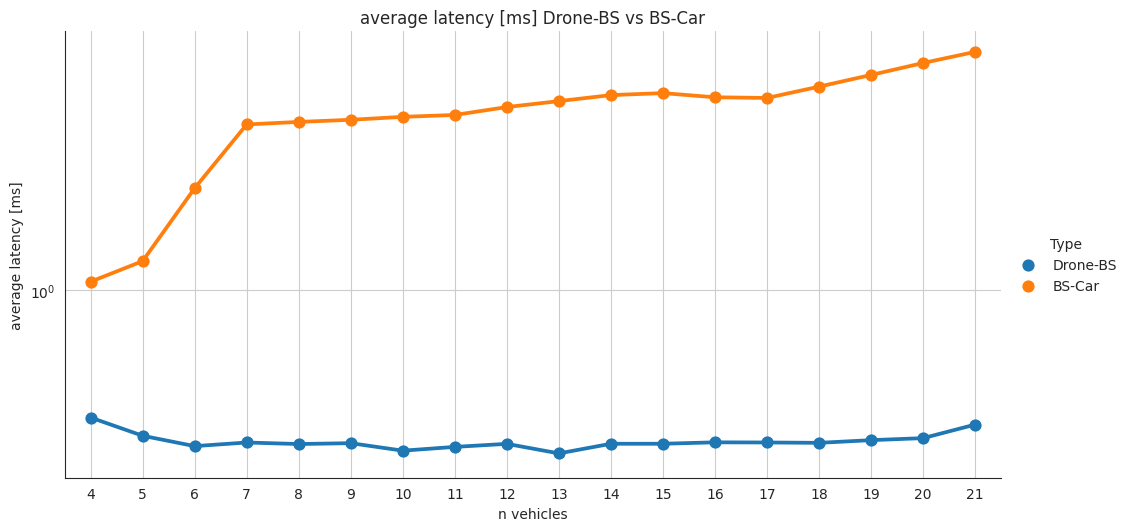}
     \caption{Average per-user latency in the uplink (UAV-\gls{BS}, or $L1$) and downlink (\gls{BS}-vehicles, or $L2$) considering the BFA scenario.}
     \label{fig:LatencyAVG4}
\end{figure}

In Fig. \ref{fig:LatencyAVG4} we plot the average per-user latency in the uplink $L_1$, i.e., the UAV-BS link, and in the downlink $L_2$, i.e., the BS-car links. 
We can see that in BFA the throughput in the uplink, where full frames are transmitted, is higher than the throughput in the downlink. Nonetheless, the latency is lower in the first link. This can be explained by considering the resource scheduling process implemented in the simulated 5G BS. As discussed in Sec.~\ref{scenimplementation}, this follows a TDD scheme, where the resources are first split between the uplink and the downlink, and then assigned to each uplink or downlink user. In this case, the UAV is the only uplink user -- thus it does not have to contend for resources with other users. Additionally, to accommodate for the analog beamforming implemented at the BS to improve the link budget with the vehicles at  \glspl{mmwave}, each symbol at the PHY layer is allocated to a single user at a time. This further deteriorates the contention performance in the downlink, and results into lower efficiency (in the resource allocation) with higher latency. Future developments can try to address this by using different scheduler implementations, frequency range (which does not require beamforming), or hybrid beamforming schemes~\cite{cuba2022hybrid}.

\textbf{\GLS{BAO} scenario. }
In this case we assume data processing at the UAV, and the transmission of only annotations (with a size proportional to the number of vehicles) in the two links. 
While the throughput is as low as 0.208 Mbit/s (0.104 Mbit/s) for 30 (15) FPS, as expected, the reliability increases with respect to BFA to 99.66\% for 30 FPS and 99.88\% for 15 FPS. 
The latency is 2.922 ms on average for both 30 and 15 FPS, regardless of the number of vehicles in the scenario. 


\textbf{Overall end-to-end comparison.}
This analysis clearly highlights that throughput should not be the only metric used to profile the performance of data dissemination systems in the context of vehicular networks. 
%
A more important metric, which is independent of the application source rate, is indeed the end-to-end reliability, which indicates (in this case) how many frames or annotations are received correctly. The throughput analysis then can provide inputs on what kind of dissemination strategy a certain wireless network (in this case, a 5G mmWave deployment) can support. 
In Fig.~\ref{fig:throughputComparison} the highest throughput is obtained in the \GLS{BFF} scenario with 30 FPS, where the UAV is sending only one copy of each frame in the first link and then the \gls{BS} is broadcasting the received information to each connected vehicle.  However, with more than 20 users the latency drastically increases (Fig.~\ref{fig:latencyComparison}) and the reliability decreases (Fig. \ref{fig:reliabilityComparison}). 
The scenarios that are transmitting only annotations (\GLS{BFA} only in the downlink, \GLS{BAO} in both uplink and downlink) have the highest reliability and the lowest latency overall. Moreover, in both scenarios, the information that each car is receiving is the same, but in one case the object detection has to be performed by the BS while in the other it has to be performed by the UAV. 
On one side, the UAV has generally more energy and computational constraints than the BS, which makes BOA more desirable. On the other side, BFA may be the only available choice in those environments lacking coverage from terrestrial infrastructures~\cite{Chaoub20216g}. Further studies on power consumption will help understand what is the final choice for UAV-to-ground communication, depending on whether it is feasible to perform object detection at the UAV.

\section{Conclusions} \label{conclusion}
In this paper we presented and evaluated different communication techniques between \gls{uav}s and ground vehicles for the dissemination of UAV sensory observations to augment vehicles' autonomous driving capabilities, based on high-capacity \gls{mmwave} links. We assessed the performance of four different communication scenarios, with applications transmitting data at 15 and 30 FPS with UDP at the transport layer.
An extensive performance evaluation based on real-world UAV data and considering ns-3 simulations showed that those configurations that transmit annotations (rather than full frames) achieve the best performance in terms of latency and reliability. For up to 21 connected users, they guarantee a latency of around 2 ms, and a reliability above 99\%. 
Our results provide a first, quantitative evaluation of the feasibility of complementing vehicle's on-board sensors with UAV data from the sky.

As part of our future work, we will combine the communication network model with an energy model that profiles the computational complexity of the object detection task on the UAV, BS, and ground vehicles hardware. 



\bibliographystyle{IEEEtran}
\bibliography{biblio.bib}

\end{document}

%% file: glossary.tex
\newacronym{3gpp}{3GPP}{3rd Generation Partnership Project}
\newacronym{4g}{4G}{4th generation}
\newacronym{5g}{5G}{5th generation}
\newacronym{6g}{6G}{6th generation}
\newacronym{5gc}{5GC}{5G Core}
\newacronym{adc}{ADC}{Analog to Digital Converter}
\newacronym{aerpaw}{AERPAW}{Aerial Experimentation and Research Platform for Advanced Wireless}
\newacronym{ai}{AI}{Artificial Intelligence}
\newacronym{aimd}{AIMD}{Additive Increase Multiplicative Decrease}
\newacronym{am}{AM}{Acknowledged Mode}
\newacronym{AMC}{AMC}{Adaptive Modulation and Coding}
\newacronym{amf}{AMF}{Access and Mobility Management Function}
\newacronym{aops}{AOPS}{Adaptive Order Prediction Scheduling}
\newacronym{api}{API}{Application Programming Interface}
\newacronym{apn}{APN}{Access Point Name}
\newacronym{ap}{AP}{Application Protocol}
\newacronym{aqm}{AQM}{Active Queue Management}
\newacronym{asn1}{ASN.1}{Abstract Syntax Notation One}
\newacronym{ausf}{AUSF}{Authentication Server Function}
\newacronym{AV}{AV}{Autonomous driving vehicle}
\newacronym{avc}{AVC}{Advanced Video Coding}
\newacronym{awgn}{AGWN}{Additive White Gaussian Noise}
\newacronym{balia}{BALIA}{Balanced Link Adaptation Algorithm}
\newacronym{BAO}{BAO}{Broadcast annotations only}
\newacronym{bbu}{BBU}{Base Band Unit}
\newacronym{bdp}{BDP}{Bandwidth-Delay Product}
\newacronym{ber}{BER}{Bit Error Rate}
\newacronym{bf}{BF}{Beamforming}
\newacronym{BFA}{BFA}{Broadcast frames and annotations}
\newacronym{BFF}{BFF}{Broadcast full frames}
\newacronym{bler}{BLER}{Block Error Rate}
\newacronym{brr}{BRR}{Bayesian Ridge Regressor}
\newacronym{BS}{BS}{base station}
\newacronym{bsr}{BSR}{Buffer Status Report}
\newacronym{bss}{BSS}{Business Support System}
\newacronym{ca}{CA}{Carrier Aggregation}
\newacronym{caas}{CaaS}{Connectivity-as-a-Service}
\newacronym{cb}{CB}{Code Block}
\newacronym{cc}{CC}{Congestion Control}
\newacronym{compc}{CC}{Component Carrier}
\newacronym{ccid}{CCID}{Congestion Control ID}
\newacronym{cco}{CC}{Carrier Component}
\newacronym{cdd}{CDD}{Cyclic Delay Diversity}
\newacronym{cdf}{CDF}{Cumulative Distribution Function}
\newacronym{cdn}{CDN}{Content Distribution Network}
\newacronym{cn}{CN}{Core Network}
\newacronym{codel}{CoDel}{Controlled Delay Management}
\newacronym{comac}{COMAC}{Converged Multi-Access and Core}
\newacronym{cord}{CORD}{Central Office Re-architected as a Datacenter}
\newacronym{cornet}{CORNET}{COgnitive Radio NETwork}
\newacronym{cosmos}{COSMOS}{Cloud Enhanced Open Software Defined Mobile Wireless Testbed for City-Scale Deployment}
\newacronym{cots}{COTS}{Commercial Off-the-Shelf}
\newacronym{cp}{CP}{Control Plane}
\newacronym{cpu}{CPU}{Central Processing Unit}
\newacronym{cqi}{CQI}{Channel Quality Information}
\newacronym{cr}{CR}{Cognitive Radio}
\newacronym{cql}{CQL}{Conservative Q-Learning}
\newacronym{cran}{CRAN}{Cloud \gls{ran}}
\newacronym{crs}{CRS}{Cell Reference Signal}
\newacronym{csi}{CSI}{Channel State Information}
\newacronym{csirs}{CSI-RS}{Channel State Information - Reference Signal}
\newacronym{cu}{CU}{Centralized Unit}
\newacronym{cucp}{CU-CP}{Centralized Unit - Control Plane}
\newacronym{cuup}{CU-UP}{Centralized Unit - User Plane}
\newacronym{d2tcp}{D$^2$TCP}{Deadline-aware Data center TCP}
\newacronym{d3}{D$^3$}{Deadline-Driven Delivery}
\newacronym{dac}{DAC}{Digital to Analog Converter}
\newacronym{dag}{DAG}{Directed Acyclic Graph}
\newacronym{das}{DAS}{Distributed Antenna System}
\newacronym{dash}{DASH}{Dynamic Adaptive Streaming over HTTP}
\newacronym{dc}{DC}{Dual Connectivity}
\newacronym{dccp}{DCCP}{Datagram Congestion Control Protocol}
\newacronym{dce}{DCE}{Direct Code Execution}
\newacronym{dci}{DCI}{Downlink Control Information}
\newacronym{dctcp}{DCTCP}{Data Center TCP}
\newacronym{dl}{DL}{Downlink}
\newacronym{dmr}{DMR}{Deadline Miss Ratio}
\newacronym{dmrs}{DMRS}{DeModulation Reference Signal}
\newacronym{drlcc}{DRL-CC}{Deep Reinforcement Learning Congestion Control}
\newacronym{drb}{DRB}{Data Radio Bearer}
\newacronym{drs}{DRS}{Discovery Reference Signal}
\newacronym{dqn}{DQN}{Deep Q-Networks}
\newacronym{du}{DU}{Distributed Unit}
\newacronym{e2e}{E2E}{end-to-end}
\newacronym{e2ap}{E2AP}{E2 Application Protocol}
\newacronym{e2sm}{E2SM}{E2 Service Model}
\newacronym{ecaas}{ECaaS}{Edge-Cloud-as-a-Service}
\newacronym{ecn}{ECN}{Explicit Congestion Notification}
\newacronym{edc}{EDC}{Edge Data Center}
\newacronym{edf}{EDF}{Earliest Deadline First}
\newacronym{embb}{eMBB}{Enhanced Mobile Broadband}
\newacronym{empower}{EMPOWER}{EMpowering transatlantic PlatfOrms for advanced WirEless Research}
\newacronym{enb}{eNB}{evolved Node Base}
\newacronym{endc}{EN-DC}{E-UTRAN-NR Dual Connectivity}
\newacronym{epc}{EPC}{Evolved Packet Core}
\newacronym{eps}{EPS}{Evolved Packet System}
\newacronym{es}{ES}{Edge Server}
\newacronym{etl}{ETL}{Extract, Transform and Load}
\newacronym{etsi}{ETSI}{European Telecommunications Standards Institute}
\newacronym[firstplural=Estimated Times of Arrival (ETAs)]{eta}{ETA}{Estimated Time of Arrival}
\newacronym{eutran}{E-UTRAN}{Evolved Universal Terrestrial Access Network}
\newacronym{faas}{FaaS}{Function-as-a-Service}
\newacronym{fapi}{FAPI}{Functional Application Platform Interface}
\newacronym{fcaps}{FCAPS}{Fault, Configuration, Accounting, Performance and Security}
\newacronym{fdd}{FDD}{Frequency Division Duplexing}
\newacronym{fdm}{FDM}{Frequency Division Multiplexing}
\newacronym{fdma}{FDMA}{Frequency Division Multiple Access}
\newacronym{fed4fire}{FED4FIRE+}{Federation 4 Future Internet Research and Experimentation Plus}
\newacronym{fir}{FIR}{Finite Impulse Response}
\newacronym{fit}{FIT}{Future \acrlong{iot}}
\newacronym{fpga}{FPGA}{Field Programmable Gate Array}
\newacronym{fr1}{FR1}{Frequency Range 1}
\newacronym{fr2}{FR2}{Frequency Range 2}
\newacronym{fs}{FS}{Fast Switching}
\newacronym{fscc}{FSCC}{Flow Sharing Congestion Control}
\newacronym{ftp}{FTP}{File Transfer Protocol}
\newacronym{fw}{FW}{Flow Window}
\newacronym{gbr}{GBR}{Guaranteed Bit Rate}
\newacronym{ge}{GE}{Gaussian Elimination}
\newacronym{gnb}{gNB}{Next Generation Node Base}
\newacronym{gop}{GOP}{Group of Pictures}
\newacronym{gpr}{GPR}{Gaussian Process Regressor}
\newacronym{gpu}{GPU}{Graphics Processing Unit}
\newacronym{gtp}{GTP}{GPRS Tunneling Protocol}
\newacronym{gtpc}{GTP-C}{GPRS Tunnelling Protocol Control Plane}
\newacronym{gtpu}{GTP-U}{GPRS Tunnelling Protocol User Plane}
\newacronym{gtpv2c}{GTPv2-C}{\gls{gtp} v2 - Control}
\newacronym{gw}{GW}{Gateway}
\newacronym{harq}{HARQ}{Hybrid Automatic Repeat reQuest}
\newacronym{hetnet}{HetNet}{Heterogeneous Network}
\newacronym{hh}{HH}{Hard Handover}
\newacronym{hol}{HOL}{Head-of-Line}
\newacronym{hqf}{HQF}{Highest-quality-first}
\newacronym{hss}{HSS}{Home Subscription Server}
\newacronym{http}{HTTP}{HyperText Transfer Protocol}
\newacronym{ia}{IA}{Initial Access}
\newacronym{iab}{IAB}{Integrated Access and Backhaul}
\newacronym{ic}{IC}{Incident Command}
\newacronym{ietf}{IETF}{Internet Engineering Task Force}
\newacronym{imsi}{IMSI}{International Mobile Subscriber Identity}
\newacronym{imt}{IMT}{International Mobile Telecommunication}
\newacronym{iot}{IoT}{Internet of Things}
\newacronym{IP}{IP}{Internet Protocol}
\newacronym{itu}{ITU}{International Telecommunication Union}
\newacronym{kpi}{KPI}{Key Performance Indicator}
\newacronym{kpm}{KPM}{Key Performance Measurement}
\newacronym{kvm}{KVM}{Kernel-based Virtual Machine}
\newacronym{los}{LOS}{Line-of-Sight}
\newacronym{ldc}{LDC}{Local Data Center}
\newacronym{lsm}{LSM}{Link-to-System Mapping}
\newacronym{lstm}{LSTM}{Long Short Term Memory}
\newacronym{lte}{LTE}{Long Term Evolution}
\newacronym{lxc}{LXC}{Linux Container}
\newacronym{m2m}{M2M}{Machine to Machine}
\newacronym{mac}{MAC}{Medium Access Control}
\newacronym{manet}{MANET}{Mobile Ad Hoc Network}
\newacronym{mano}{MANO}{Management and Orchestration}
\newacronym{mbr}{MBR}{Maximum Bit Rate}
\newacronym{mc}{MC}{Multi-Connectivity}
\newacronym{mcc}{MCC}{Mobile Cloud Computing}
\newacronym{mchem}{MCHEM}{Massive Channel Emulator}
\newacronym{mcs}{MCS}{Modulation and Coding Scheme}
\newacronym{mec}{MEC}{Multi-access Edge Computing}
\newacronym{mec2}{MEC}{Mobile Edge Cloud}
\newacronym{mfc}{MFC}{Mobile Fog Computing}
\newacronym{MFF}{MFF}{Multiple full frames}
\newacronym{mi}{MI}{Mutual Information}
\newacronym{mib}{MIB}{Master Information Block}
\newacronym{miesm}{MIESM}{Mutual Information Based Effective SINR}
\newacronym{mimo}{MIMO}{Multiple Input Multiple Output}
\newacronym{ml}{ML}{Machine Learning}
\newacronym{mlr}{MLR}{Maximum-local-rate}
\newacronym[plural=\gls{mme}s,firstplural=Mobility Management Entities (MMEs)]{mme}{MME}{Mobility Management Entity}
\newacronym{mmtc}{mMTC}{Massive Machine-Type Communications}
\newacronym{mmwave}{mmWave}{millimeter wave}
\newacronym{mpdccp}{MP-DCCP}{Multipath Datagram Congestion Control Protocol}
\newacronym{mptcp}{MPTCP}{Multipath TCP}
\newacronym{mr}{MR}{Maximum Rate}
\newacronym{mrdc}{MR-DC}{Multi \gls{rat} \gls{dc}}
\newacronym{mse}{MSE}{Mean Square Error}
\newacronym{mdp}{MDP}{Markov Decision Process}
\newacronym{mss}{MSS}{Maximum Segment Size}
\newacronym{mt}{MT}{Mobile Termination}
\newacronym{mtd}{MTD}{Machine-Type Device}
\newacronym{mtu}{MTU}{Maximum Transmission Unit}
\newacronym{mumimo}{MU-MIMO}{Multi-user \gls{mimo}}
\newacronym{mvno}{MVNO}{Mobile Virtual Network Operator}
\newacronym{nalu}{NALU}{Network Abstraction Layer Unit}
\newacronym{nas}{NAS}{Non-Access Stratum}
\newacronym{ngran}{NG-RAN}{Next Generation - \gls{ran}}
\newacronym{ns-3}{ns-3}{Network Simulator 3}
\newacronym{nbiot}{NB-IoT}{Narrow Band IoT}
\newacronym{nf}{NF}{Network Function}
\newacronym{nfv}{NFV}{Network Function Virtualization}
\newacronym{nfvi}{NFVI}{Network Function Virtualization Infrastructure}
\newacronym{nic}{NIC}{Network Interface Card}
\newacronym{nlos}{NLOS}{Non-Line-of-Sight}
\newacronym{now}{NOW}{Non Overlapping Window}
\newacronym{nsm}{NSM}{Network Service Mesh}
\newacronym{NR}{NR}{New Radio}
\newacronym{nrf}{NRF}{Network Repository Function}
\newacronym{nsa}{NSA}{Non Stand Alone}
\newacronym{nse}{NSE}{Network Slicing Engine}
\newacronym{nssf}{NSSF}{Network Slice Selection Function}
\newacronym{o2i}{O2I}{Outdoor to Indoor}
\newacronym{oai}{OAI}{OpenAirInterface}
\newacronym{oaicn}{OAI-CN}{\gls{oai} \acrlong{cn}}
\newacronym{oairan}{OAI-RAN}{\acrlong{oai} \acrlong{ran}}
\newacronym{oam}{OAM}{Operations, Administration and Maintenance}
\newacronym{OFDM}{OFDM}{Orthogonal Frequency Division Multiplexing}
\newacronym{olia}{OLIA}{Opportunistic Linked Increase Algorithm}
\newacronym{omec}{OMEC}{Open Mobile Evolved Core}
\newacronym{onap}{ONAP}{Open Network Automation Platform}
\newacronym{onf}{ONF}{Open Networking Foundation}
\newacronym{onos}{ONOS}{Open Networking Operating System}
\newacronym{oom}{OOM}{\gls{onap} Operations Manager}
\newacronym{opnfv}{OPNFV}{Open Platform for \gls{nfv}}
\newacronym{oran}{O-RAN}{Open \gls{ran}}
\newacronym{orbit}{ORBIT}{Open-Access Research Testbed for Next-Generation Wireless Networks}
\newacronym{os}{OS}{Operating System}
\newacronym{oss}{OSS}{Operations Support System}
\newacronym{pa}{PA}{Position-aware}
\newacronym{pase}{PASE}{Prioritization, Arbitration, and Self-adjusting Endpoints}
\newacronym{pawr}{PAWR}{Platforms for Advanced Wireless Research}
\newacronym{pbch}{PBCH}{Physical Broadcast Channel}
\newacronym{pcef}{PCEF}{Policy and Charging Enforcement Function}
\newacronym{pcfich}{PCFICH}{Physical Control Format Indicator Channel}
\newacronym{pcrf}{PCRF}{Policy and Charging Rules Function}
\newacronym{pdcch}{PDCCH}{Physical Downlink Control Channel}
\newacronym{pdcp}{PDCP}{Packet Data Convergence Protocol}
\newacronym{pdcpc}{PDCP-C}{Packet Data Convergence Protocol - Control Plane}
\newacronym{pdcpu}{PDCP-U}{Packet Data Convergence Protocol - User Plane}
\newacronym{pdsch}{PDSCH}{Physical Downlink Shared Channel}
\newacronym{pdu}{PDU}{Packet Data Unit}
\newacronym{pf}{PF}{Proportional Fair}
\newacronym{pgw}{PGW}{Packet Gateway}
\newacronym{phich}{PHICH}{Physical Hybrid ARQ Indicator Channel}
\newacronym{phy}{PHY}{Physical}
\newacronym{phyu}{PHY-U}{Upper Physical}
\newacronym{phyl}{PHY-L}{Lower Physical}
\newacronym{pmch}{PMCH}{Physical Multicast Channel}
\newacronym{pmi}{PMI}{Precoding Matrix Indicators}
\newacronym{powder}{POWDER}{Platform for Open Wireless Data-driven Experimental Research}
\newacronym{ppo}{PPO}{Proximal Policy Optimization}
\newacronym{ppp}{PPP}{Poisson Point Process}
\newacronym{prach}{PRACH}{Physical Random Access Channel}
\newacronym{prb}{PRB}{Physical Resource Block}
\newacronym{psnr}{PSNR}{Peak Signal to Noise Ratio}
\newacronym{pss}{PSS}{Primary Synchronization Signal}
\newacronym{pucch}{PUCCH}{Physical Uplink Control Channel}
\newacronym{pusch}{PUSCH}{Physical Uplink Shared Channel}
\newacronym{qam}{QAM}{Quadrature Amplitude Modulation}
\newacronym{qci}{QCI}{\gls{qos} Class Identifier}
\newacronym{qoe}{QoE}{Quality of Experience}
\newacronym{qos}{QoS}{Quality of Service}
\newacronym{quic}{QUIC}{Quick UDP Internet Connections}
\newacronym{rach}{RACH}{Random Access Channel}
\newacronym{ran}{RAN}{Radio Access Network}
\newacronym[firstplural=Radio Access Technologies (RATs)]{rat}{RAT}{Radio Access Technology}
\newacronym{rcn}{RCN}{Research Coordination Network}
\newacronym{rec}{REC}{Radio Edge Cloud}
\newacronym{red}{RED}{Random Early Detection}
\newacronym{rem}{REM}{Random Ensemble Mixture}
\newacronym{renew}{RENEW}{Reconfigurable Eco-system for Next-generation End-to-end Wireless}
\newacronym{rf}{RF}{Radio Frequency}
\newacronym{rfc}{RFC}{Request for Comments}
\newacronym{rfr}{RFR}{Random Forest Regressor}
\newacronym{ric}{RIC}{\gls{ran} Intelligent Controller}
\newacronym{rlc}{RLC}{Radio Link Control}
\newacronym{rl}{RL}{Reinforcement Learning}
\newacronym{rlf}{RLF}{Radio Link Failure}
\newacronym{rlnc}{RLNC}{Random Linear Network Coding}
\newacronym{rmr}{RMR}{RIC Message Routing}
\newacronym{rmse}{RMSE}{Root Mean Squared Error}
\newacronym{rnis}{RNIS}{Radio Network Information Service}
\newacronym{rnib}{RNIB}{Radio Network Information Base}
\newacronym{RNTI}{RNTI}{Radio Network Temporary Identifier}
\newacronym{rr}{RR}{Round Robin}
\newacronym{rrc}{RRC}{Radio Resource Control}
\newacronym{rrm}{RRM}{Radio Resource Management}
\newacronym{rru}{RRU}{Remote Radio Unit}
\newacronym{rs}{RS}{Remote Server}
\newacronym{rsrp}{RSRP}{Reference Signal Received Power}
\newacronym{rsrq}{RSRQ}{Reference Signal Received Quality}
\newacronym{rss}{RSS}{Received Signal Strength}
\newacronym{rssi}{RSSI}{Received Signal Strength Indicator}
\newacronym{rtt}{RTT}{Round Trip Time}
\newacronym{ru}{RU}{Radio Unit}
\newacronym{rw}{RW}{Receive Window}
\newacronym{rx}{RX}{Receiver}
\newacronym{s1ap}{S1AP}{S1 Application Protocol}
\newacronym{sa}{SA}{standalone}
\newacronym{sack}{SACK}{Selective Acknowledgment}
\newacronym{SAE}{SAE}{Society of Automotive Engineers}
\newacronym{sap}{SAP}{Service Access Point}
\newacronym{sc2}{SC2}{Spectrum Collaboration Challenge}
\newacronym{scef}{SCEF}{Service Capability Exposure Function}
\newacronym{sch}{SCH}{Secondary Cell Handover}
\newacronym{scoot}{SCOOT}{Split Cycle Offset Optimization Technique}
\newacronym{sctp}{SCTP}{Stream Control Transmission Protocol}
\newacronym{sdap}{SDAP}{Service Data Adaptation Protocol}
\newacronym{sdk}{SDK}{Software Development Kit}
\newacronym{sdm}{SDM}{Space Division Multiplexing}
\newacronym{sdma}{SDMA}{Spatial Division Multiple Access}
\newacronym{sdn}{SDN}{Software-defined Networking}
\newacronym{sdr}{SDR}{Software-defined Radio}
\newacronym{seba}{SEBA}{SDN-Enabled Broadband Access}
\newacronym{sgsn}{SGSN}{Serving GPRS Support Node}
\newacronym{sgw}{SGW}{Serving Gateway}
\newacronym{si}{SI}{Study Item}
\newacronym{sib}{SIB}{Secondary Information Block}
\newacronym{sinr}{SINR}{Signal to Interference plus Noise Ratio}
\newacronym{sip}{SIP}{Session Initiation Protocol}
\newacronym{siso}{SISO}{Single Input, Single Output}
\newacronym{sla}{SLA}{Service Level Agreement}
\newacronym{sm}{SM}{Service Model}
\newacronym{smf}{SMF}{Session Management Function}
\newacronym{smo}{SMO}{Service Management and Orchestration}
\newacronym{sms}{SMS}{Short Message Service}
\newacronym{smsgmsc}{SMS-GMSC}{\gls{sms}-Gateway}
\newacronym{snr}{SNR}{Signal-to-Noise-Ratio}
\newacronym{son}{SON}{Self-Organizing Network}
\newacronym{sptcp}{SPTCP}{Single Path TCP}
\newacronym{srb}{SRB}{Service Radio Bearer}
\newacronym{srn}{SRN}{Standard Radio Node}
\newacronym{srs}{SRS}{Sounding Reference Signal}
\newacronym{ss}{SS}{Synchronization Signal}
\newacronym{sss}{SSS}{Secondary Synchronization Signal}
\newacronym{st}{ST}{Spanning Tree}
\newacronym{svc}{SVC}{Scalable Video Coding}
\newacronym{tb}{TB}{Transport Block}
\newacronym{tcp}{TCP}{Transmission Control Protocol}
\newacronym{TDD}{TDD}{Time Division Duplexing}
\newacronym{tdm}{TDM}{Time Division Multiplexing}
\newacronym{tdma}{TDMA}{Time Division Multiple Access}
\newacronym{tfl}{TfL}{Transport for London}
\newacronym{tfrc}{TFRC}{TCP-Friendly Rate Control}
\newacronym{tft}{TFT}{Traffic Flow Template}
\newacronym{tgen}{TGEN}{Traffic Generator}
\newacronym{tip}{TIP}{Telecom Infra Project}
\newacronym{tm}{TM}{Transparent Mode}
\newacronym{to}{TO}{Telco Operator}
\newacronym{tr}{TR}{Technical Report}
\newacronym{trp}{TRP}{Transmitter Receiver Pair}
\newacronym{ts}{TS}{Traffic Steering}
\newacronym{tti}{TTI}{Transmission Time Interval}
\newacronym{ttt}{TTT}{Time-to-Trigger}
\newacronym{tx}{TX}{Transmitter}
\newacronym{uas}{UAS}{Unmanned Aerial System}
\newacronym{uav}{UAV}{Unmanned Aerial Vehicle}
\newacronym{udm}{UDM}{Unified Data Management}
\newacronym{udp}{UDP}{User Datagram Protocol}
\newacronym{udr}{UDR}{Unified Data Repository}
\newacronym{uhd}{UHD}{\gls{usrp} Hardware Driver}
\newacronym{ul}{UL}{Uplink}
\newacronym{um}{UM}{Unacknowledged Mode}
\newacronym{uml}{UML}{Unified Modeling Language}
\newacronym{upa}{UPA}{Uniform Planar Array}
\newacronym{upf}{UPF}{User Plane Function}
\newacronym{urllc}{URLLC}{Ultra Reliable and Low Latency Communications}
\newacronym{usa}{U.S.}{United States}
\newacronym{usim}{USIM}{Universal Subscriber Identity Module}
\newacronym{usrp}{USRP}{Universal Software Radio Peripheral}
\newacronym{utc}{UTC}{Urban Traffic Control}
\newacronym{vim}{VIM}{Virtualization Infrastructure Manager}
\newacronym{vm}{VM}{Virtual Machine}
\newacronym{vnf}{vNF}{Virtualized Network Function}
\newacronym{volte}{VoLTE}{Voice over \gls{lte}}
\newacronym{voltha}{VOLTHA}{Virtual OLT HArdware Abstraction}
\newacronym{vr}{VR}{Virtual Reality}
\newacronym{vran}{vRAN}{Virtualized \gls{ran}}
\newacronym{vss}{VSS}{Video Streaming Server}
\newacronym{v2x}{V2X}{Vehicle-to-everything}
\newacronym{wbf}{WBF}{Wired Bias Function}
\newacronym{wf}{WF}{Waterfilling}
\newacronym{wlan}{WLAN}{Wireless Local Area Network}
\newacronym{osm}{OSM}{Open Source \gls{nfv} Management and Orchestration}
\newacronym{pnf}{PNF}{Physical Network Function}
\newacronym{drl}{DRL}{Deep Reinforcement Learning}
\newacronym{mtc}{MTC}{Machine-type Communications}
\newacronym{osc}{OSC}{O-RAN Software Community}
\newacronym{rc}{RC}{RAN Control}
\newacronym{ar}{AR}{Augmented Reality}
\newacronym{daps}{DAPS}{Dual Active Protocol Stack}
\newacronym{nib}{NIB}{Network Information Base}
\newacronym{ue}{UE}{User Equipment}

%% file: images/ThroughputEndToEndUSerX.tex
\begin{tikzpicture} [every mark/.append style={mark size=1pt}]
\pgfplotsset{every tick label/.append style={font=\scriptsize}}

\definecolor{color0}{rgb}{0.12156862745098,0.466666666666667,0.705882352941177}
\definecolor{color1}{rgb}{1,0.498039215686275,0.0549019607843137}
\definecolor{color2}{rgb}{0.172549019607843,0.627450980392157,0.172549019607843}
\definecolor{color3}{rgb}{0.83921568627451,0.152941176470588,0.156862745098039}
\definecolor{color4}{rgb}{0.580392156862745,0.403921568627451,0.741176470588235}
\definecolor{color5}{rgb}{0.549019607843137,0.337254901960784,0.294117647058824}
\definecolor{color6}{rgb}{0.890196078431372,0.466666666666667,0.76078431372549}

\begin{axis}[
width=\fwidth,
height=\fheight,
at={(0\fwidth,0\fheight)},
scale only axis,
axis line style={white!15!black},
x grid style={white!80!black},
xlabel={Number of vehicles},
xmajorgrids,
xmajorticks=true,
xmin=4, xmax=21,
xlabel style={font=\scriptsize\color{white!15!black}},
ylabel style={font=\scriptsize\color{white!15!black}},
xtick style={color=white!15!black},
y grid style={white!80!black},
ylabel={Throughput [Mbit/s]},
ymajorgrids,
ymajorticks=true,
ymin=0, ymax=41.173356505,
ytick style={color=white!15!black},
legend style={legend cell align=left, align=left, draw=white!15!black,
anchor=south west, font=\scriptsize, align=left, at={(0.1,1.2)}},
legend columns=8,
]

\addplot [line width=1pt, mark options={solid, scale=0.5}, color0, mark=*]
table {%
4 39.198753
5 38.98823928
6 37.439134
7 37.0179041142857
8 35.71012605
9 36.479178
10 34.14218052
11 33.4230243272727
12 29.9805557
13 27.7300558153846
14 24.6885594
15 23.91293168
16 22.4394696
17 21.0329232
18 19.5935483333333
19 18.4682789052632
20 17.84520192
21 16.7707525714286
};
\addlegendentry{MFF, 30 fps};

\addplot [line width=1pt, mark options={solid, scale=0.5}, color1, mark=o, densely dashed]
table {%
4 19.5993357
5 19.49410944
6 18.7192032
7 18.5127202285714
8 17.85533715
9 18.2559078666667
10 17.07306804
11 17.5322496
12 17.8265247
13 17.1566652
14 17.8838756571429
15 17.69826752
16 18.1079325
17 18.1130856
18 18.0151244
19 18.1787846526316
20 17.89793592
21 16.9539241714286
};
\addlegendentry{MFF, 15fps};

\addplot [line width=1pt, mark options={solid, scale=0.5}, color2, mark=x]
table {%
4 39.1989213
5 38.9886228
6 37.4396848
7 37.0251752571429
8 35.711424
9 36.5122305333333
10 34.14758652
11 35.0645882181818
12 35.6513681
13 34.3135736307692
14 35.5267573714286
15 35.37873944
16 36.2000652
17 35.9915832
18 35.9326552
19 35.7796115368421
20 37.04721498
21 36.0148604
};
\addlegendentry{BFF, 30fps};

\addplot [line width=1pt, mark options={solid, scale=0.5}, color3, mark=triangle, densely dashed]
table {%
4 19.5993408
5 19.4940768
6 18.7195296
7 18.5127464571429
8 17.8554825
9 18.2559668
10 17.0732802
11 17.5320715636364
12 17.8267236
13 17.1568284
14 17.8843492285714
15 17.6986252
16 18.1087434
17 18.1138512
18 18.0164266
19 18.2084977894737
20 18.96453054
21 19.9541366285714
};
\addlegendentry{BFF, 15fps};

\addplot [line width=1pt, mark options={solid, scale=0.5}, color4, mark=diamond]
table {%
4 0.0447790933333333
5 0.0542336
6 0.06371968
7 0.0730578133333333
8 0.0825767466666667
9 0.0922608
10 0.101705706666667
11 0.11064704
12 0.120198613333333
13 0.12981696
14 0.1387776
15 0.148533333333333
16 0.1587488
17 0.167458133333333
18 0.17713152
19 0.186487466666667
20 0.19617792
21 0.205647146666667
};
\addlegendentry{BFA, 30fps};

\addplot [line width=1pt, mark options={solid, scale=0.5}, color5, mark=square, densely dashed]
table {%
4 0.02237952
5 0.0271168
6 0.03181696
7 0.0365125333333333
8 0.0412698666666667
9 0.0460593066666667
10 0.05083008
11 0.0552738133333333
12 0.06007232
13 0.06487936
14 0.0693888
15 0.0742
16 0.0793034666666667
17 0.0836539733333333
18 0.0884864
19 0.0932437333333333
20 0.09808896
21 0.102639466666667
};
\addlegendentry{BFA, 15fps};

\addplot [line width=1pt, mark options={solid, scale=0.5}, color6, mark=otimes]
table {%
4 0.0454208
5 0.0550848
6 0.0647488
7 0.0741712
8 0.0838352
9 0.0934992
10 0.1031632
11 0.1125856
12 0.1222496
13 0.1319136
14 0.141336
15 0.151
16 0.160664
17 0.1700864
18 0.1797504
19 0.1894144
20 0.1990784
21 0.2085008
};
\addlegendentry{BAO, 30fps};

\addplot [line width=1pt, mark options={solid, scale=0.5}, white!49.8039215686275!black, mark=star, densely dashed]
table {%
4 0.0227605333333333
5 0.0276032
6 0.0324458666666667
7 0.0371674666666667
8 0.0419916266666667
9 0.0468528
10 0.0516783866666667
11 0.0564170666666667
12 0.0612327466666667
13 0.06607328
14 0.070824
15 0.0756083333333333
16 0.0805093333333333
17 0.0852309333333333
18 0.0900736
19 0.0948849066666667
20 0.09967104
21 0.104480533333333
};
\addlegendentry{BAO, 15fps};

\end{axis}

\end{tikzpicture}

%% file: images/avg_latencyEndToEndUSerX.tex
\begin{tikzpicture}[every mark/.append style={mark size=1pt}]
\pgfplotsset{every tick label/.append style={font=\scriptsize}}

\definecolor{color0}{rgb}{0.12156862745098,0.466666666666667,0.705882352941177}
\definecolor{color1}{rgb}{1,0.498039215686275,0.0549019607843137}
\definecolor{color2}{rgb}{0.172549019607843,0.627450980392157,0.172549019607843}
\definecolor{color3}{rgb}{0.83921568627451,0.152941176470588,0.156862745098039}
\definecolor{color4}{rgb}{0.580392156862745,0.403921568627451,0.741176470588235}
\definecolor{color5}{rgb}{0.549019607843137,0.337254901960784,0.294117647058824}
\definecolor{color6}{rgb}{0.890196078431372,0.466666666666667,0.76078431372549}

\begin{axis}[
width=\fwidth,
height=\fheight,
at={(0\fwidth,0\fheight)},
scale only axis,
axis line style={white!15!black},
x grid style={white!80!black},
xlabel={Number of vehicles},
xmajorgrids,
xmajorticks=true,
xmin=4, xmax=21,
xtick style={color=white!15!black},
xlabel style={font=\scriptsize\color{white!15!black}},
ylabel style={font=\scriptsize\color{white!15!black}},
y grid style={white!80!black},
ylabel={Average latency [ms]},
ymajorgrids,
ymajorticks=true,
ymin=0, ymax=694.548472644665,
ytick style={color=white!15!black},
legend style={legend cell align=left, align=left, draw=white!15!black,
anchor=south, font=\scriptsize, align=left, at={(0.5,1.05)}},
legend columns=8,
]

\addplot [line width=1pt, mark options={solid, scale=0.5}, color0, mark=*]
table {%
4 1.60776121573087
5 1.69537053146175
6 1.83151509291667
7 1.9978582444555
8 2.1004554585724
9 2.68677893604281
10 2.79074290136612
11 196.772936924305
12 220.790202842623
13 222.60053349938
14 235.574305530172
15 228.052956560464
16 227.42505758265
17 229.150425445998
18 231.04855409745
19 233.532901325712
20 230.994421888661
21 233.179784878415
};

\addplot [line width=1pt, mark options={solid, scale=0.5}, color1, mark=o, densely dashed]
table {%
4 1.78713333021175
5 1.81790463754098
6 1.89039976939891
7 1.91468478906128
8 1.94287163814891
9 1.90088579262295
10 2.02330824535519
11 1.98682916844262
12 1.98563914955601
13 1.99717695937369
14 2.24298629986339
15 2.24035337095628
16 2.69359335204918
17 2.93104643381148
18 3.02371471297814
19 14.9026806319924
20 202.792247573224
21 222.814093279567
};

\addplot [line width=1pt, mark options={solid, scale=0.5}, color2, mark=x]
table {%
4 1.54277638231557
5 1.52475787564208
6 1.58766874163934
7 1.69751710432279
8 1.64385604034836
9 1.72729015728142
10 1.69481711247268
11 1.78902263203925
12 1.83600551940574
13 1.84356420226093
14 2.03836741722482
15 2.04486413260929
16 2.12191391404372
17 2.2712517434603
18 2.24445491229736
19 2.50682189495614
20 2.79761989829235
21 637.1414585851
};

\addplot [line width=1pt, mark options={solid, scale=0.5}, color3, mark=triangle, densely dashed]
table {%
4 1.75618514178962
5 1.76472954638661
6 1.79075008306011
7 1.91566532418033
8 1.9296116653347
9 1.9169679034153
10 1.94204790030055
11 1.90511235628415
12 1.84658992815574
13 1.92109188650116
14 1.94141690032494
15 1.98838439439891
16 1.99598294739583
17 2.01519227035519
18 1.99441409749089
19 2.07219076133161
20 2.10489311723361
21 2.08506837217896
};

\addplot [line width=1pt, mark options={solid, scale=0.5}, color4, mark=diamond]
table {%
4 1.74496564928962
5 1.76694022
6 1.96696611284153
7 2.19969684533958
8 2.20686921486339
9 2.21660204795993
10 2.21535274327869
11 2.22918128261302
12 2.26582599982923
13 2.27366221121375
14 2.31475511780933
15 2.32299629928962
16 2.30820985724044
17 2.30513223336548
18 2.3518522629326
19 2.40663986425942
20 2.4631049307582
21 2.53781796907885
};

\addplot [line width=1pt, mark options={solid, scale=0.5}, color5, mark=square, densely dashed]
table {%
4 1.98456105420765
5 2.07338190038251
6 2.29128394699454
7 2.53708072404372
8 2.54816738852459
9 2.55497196703097
10 2.56281368404372
11 2.57350081977149
12 2.60774134788251
13 2.62307912581915
14 2.65920761055816
15 2.67614644747268
16 2.66016510191257
17 2.64749117097396
18 2.67398664517304
19 2.72195676393443
20 2.76723760054645
21 2.75533808430133
};

\addplot [line width=1pt, mark options={solid, scale=0.5}, color6, mark=otimes]
table {%
4 1.89618613114754
5 1.98950650819672
6 2.15981968852459
7 2.38713383138173
8 2.42374680327869
9 2.67865863387978
10 2.68737754098361
11 2.69639473919523
12 2.70565607581967
13 2.73512183795712
14 2.76144706088993
15 2.77804875273224
16 2.79356131147541
17 2.77851878495661
18 2.77148464480874
19 2.82026313201035
20 2.86854770491803
21 2.92201353629977
};

\addplot [line width=1pt, mark options={solid, scale=0.5}, white!49.8039215686275!black, mark=star, densely dashed]
table {%
4 1.89618613114754
5 1.98950650819672
6 2.15981968852459
7 2.38713383138173
8 2.42373686653005
9 2.67865863387978
10 2.68736589745902
11 2.69639473919523
12 2.70555831405396
13 2.73517506456494
14 2.7613906440281
15 2.77806134856557
16 2.79356131147541
17 2.77851878495661
18 2.77148464480874
19 2.82025168178027
20 2.8685241338388
21 2.92201353629977
};

\end{axis}

\end{tikzpicture}

%% file: images/Reliability_pckt_EndToEndUSerX.tex
\begin{tikzpicture} [every mark/.append style={mark size=1pt}]
\pgfplotsset{every tick label/.append style={font=\scriptsize}}

\definecolor{color0}{rgb}{0.12156862745098,0.466666666666667,0.705882352941177}
\definecolor{color1}{rgb}{1,0.498039215686275,0.0549019607843137}
\definecolor{color2}{rgb}{0.172549019607843,0.627450980392157,0.172549019607843}
\definecolor{color3}{rgb}{0.83921568627451,0.152941176470588,0.156862745098039}
\definecolor{color4}{rgb}{0.580392156862745,0.403921568627451,0.741176470588235}
\definecolor{color5}{rgb}{0.549019607843137,0.337254901960784,0.294117647058824}
\definecolor{color6}{rgb}{0.890196078431372,0.466666666666667,0.76078431372549}

\begin{axis}[
width=\fwidth,
height=\fheight,
at={(0\fwidth,0\fheight)},
scale only axis,
axis line style={white!15!black},
x grid style={white!80!black},
xlabel={Number of vehicles},
xlabel style={font=\scriptsize\color{white!15!black}},
ylabel style={font=\scriptsize\color{white!15!black}},
xmajorgrids,
xmajorticks=true,
xmin=4, xmax=21,
xtick style={color=white!15!black},
y grid style={white!80!black},
ylabel={Received packets [\%]},
ymajorgrids,
ymajorticks=true,
ymin=40, ymax=100,
ytick style={color=white!15!black}
]

\addplot [line width=0.5pt, color0, mark=*]
table {%
4 99.7487209479648
5 99.7480298676823
6 99.7476172568704
7 99.7281107539666
8 99.7439313074336
9 99.6563308250883
10 99.7318836670153
11 95.0791145462148
12 83.8785077550279
13 80.7261919657188
14 68.8485934583843
15 67.3876216830465
16 61.8001249525379
17 57.9102979982463
18 54.2486885966077
19 50.5850296084303
20 46.9296998274943
21 41.9228732627832
};

\addplot [line width=1pt, mark options={solid, scale=0.5}, color1, mark=o, densely dashed]
table {%
4 99.7485121162189
5 99.74797635274
6 99.7456804424965
7 99.7484097248032
8 99.7454615040773
9 99.7454556892292
10 99.7433419841393
11 99.746755351527
12 99.7469731251481
13 99.7450104915286
14 99.7429849144339
15 99.7428925627648
16 99.7413956780108
17 99.7409957072798
18 99.7388126903409
19 99.5843373798951
20 94.1366424182307
21 84.7612843198155
};

\addplot [line width=1pt, mark options={solid, scale=0.5}, color2, mark=x]
table {%
4 99.7491486609422
5 99.7490104693234
6 99.7490864498702
7 99.7476933286587
8 99.7475651505673
9 99.7465859868692
10 99.7475773563047
11 99.7470044607923
12 99.7422659599772
13 99.7457078395647
14 99.0709823225995
15 99.6927582368722
16 99.6978928535095
17 99.09505267049
18 99.4692386117694
19 98.0013065252108
20 97.4182327338826
21 90.0284383581424
};

\addplot [line width=1pt, mark options={solid, scale=0.5}, color3, mark=triangle, densely dashed]
table {%
4 99.7485380154399
5 99.7478092725499
6 99.7474206606865
7 99.7485511728976
8 99.7462832574295
9 99.7457758847507
10 99.7445681110511
11 99.7457415939774
12 99.748082479287
13 99.7459623045389
14 99.7456244613847
15 99.7449102329141
16 99.7458603158031
17 99.7452132628444
18 99.7460181044341
19 99.7471068195341
20 99.7442862842079
21 99.7441355700542
};

\addplot [line width=1pt, mark options={solid, scale=0.5}, color4, mark=diamond]
table {%
4 98.252
5 98.12
6 98.076
7 98.164
8 98.164
9 98.34
10 98.252
11 97.944
12 97.988
13 98.076
14 97.856
15 98.032
16 98.472
17 98.12
18 98.208
19 98.12
20 98.208
21 98.296
};


\addplot [line width=1pt, mark options={solid, scale=0.5}, color5, mark=square, densely dashed]
table {%
4 98.208
5 98.12
6 97.944
7 98.12
8 98.12
9 98.1884444444445
10 98.208
11 97.856
12 97.944
13 98.032
14 97.856
15 97.944
16 98.384
17 98.032
18 98.12
19 98.12
20 98.208
21 98.12
};

\addplot [line width=1pt, mark options={solid, scale=0.5}, color6, mark=otimes]
table {%
4 99.66
5 99.66
6 99.66
7 99.66
8 99.66
9 99.66
10 99.66
11 99.66
12 99.66
13 99.66
14 99.66
15 99.66
16 99.66
17 99.66
18 99.66
19 99.66
20 99.66
21 99.66
};

\addplot [line width=1pt, mark options={solid, scale=0.5}, white!49.8039215686275!black, mark=star, densely dashed]
table {%
4 99.88
5 99.88
6 99.88
7 99.88
8 99.836
9 99.88
10 99.847
11 99.88
12 99.836
13 99.836
14 99.88
15 99.803
16 99.88
17 99.88
18 99.88
19 99.847
20 99.792
21 99.88
};

\end{axis}

\end{tikzpicture}

%% file: images/FrameSize.tex
\begin{tikzpicture}[every mark/.append style={mark size=1pt}]
\pgfplotsset{every tick label/.append style={font=\scriptsize}}

\definecolor{color0}{rgb}{0.12156862745098,0.466666666666667,0.705882352941177}

\begin{axis}[
width=\fwidth,
height=\fheight,
at={(0\fwidth,0\fheight)},
scale only axis,
axis line style={white!15!black},
x grid style={white!80!black},
xlabel={Number of vehicles},
xmajorgrids,
xmajorticks=true,
x label style={font=\footnotesize,at={(axis description cs:0.5,-0.13)},anchor=north},
y label style={font=\footnotesize,at={(axis description cs:-0.08,0.5)},anchor=south},
xmin=4, xmax=21,
xtick style={color=white!15!black},
y grid style={white!80!black},
ylabel={Frame size [byte]},
ymajorgrids,
ymajorticks=true,
ymin=134100.12, ymax=166045.88,
ytick style={color=white!15!black}
]
table{%
x  y  draw  fill
0 160529 31,119,180 31,119,180
1 159668 31,119,180 31,119,180
2 153324.6 31,119,180 31,119,180
3 151629.2 31,119,180 31,119,180
4 146249.2 31,119,180 31,119,180
5 149530.2 31,119,180 31,119,180
6 139844.8 31,119,180 31,119,180
7 143601 31,119,180 31,119,180
8 146011 31,119,180 31,119,180
9 140527.2 31,119,180 31,119,180
10 146486.6 31,119,180 31,119,180
11 144966.4 31,119,180 31,119,180
12 148324.2 31,119,180 31,119,180
13 148367 31,119,180 31,119,180
14 147567.8 31,119,180 31,119,180
15 149139.4 31,119,180 31,119,180
16 155336.2 31,119,180 31,119,180
17 163442.2 31,119,180 31,119,180
};
\addplot [line width=0.5pt, color0, mark=*]
table {%
4 160529
5 159668
6 153324.6
7 151629.2
8 146249.2
9 149530.2
10 139844.8
11 143601
12 146011
13 140527.2
14 146486.6
15 144966.4
16 148324.2
17 148367
18 147567.8
19 149139.4
20 155336.2
21 163442.2
};
\addplot [line width=0.5pt, color0]
table {%
4 160402.2
4 160702.8
};
\addplot [line width=0.5pt, color0]
table {%
5 159571.59
5 159774.2
};
\addplot [line width=0.5pt, color0]
table {%
6 153060.2
6 153532.4
};
\addplot [line width=0.5pt, color0]
table {%
7 151433.4
7 151865.05
};
\addplot [line width=0.5pt, color0]
table {%
8 145241.6
8 147188.6
};
\addplot [line width=0.5pt, color0]
table {%
9 149240.4
9 149734.4
};
\addplot [line width=0.5pt, color0]
table {%
10 136945.6
10 141508.2
};
\addplot [line width=0.5pt, color0]
table {%
11 143091.6
11 143997.2
};
\addplot [line width=0.5pt, color0]
table {%
12 145461.8
12 146585.555
};
\addplot [line width=0.5pt, color0]
table {%
13 135552.2
13 144530.2
};
\addplot [line width=0.5pt, color0]
table {%
14 145890.4
14 147082.8
};
\addplot [line width=0.5pt, color0]
table {%
15 143832.4
15 145768.8
};
\addplot [line width=0.5pt, color0]
table {%
16 148217.6
16 148429.4
};
\addplot [line width=0.5pt, color0]
table {%
17 148195.4
17 148532.8
};
\addplot [line width=0.5pt, color0]
table {%
18 145550.6
18 148658.57
};
\addplot [line width=0.5pt, color0]
table {%
19 149097.965
19 149175.6
};
\addplot [line width=0.5pt, color0]
table {%
20 154661
20 155950.74
};
\addplot [line width=0.5pt, color0]
table {%
21 161325.945
21 164593.8
};
\end{axis}

\end{tikzpicture}

%% file: images/avg_latencyDrone-BSvsBS-Car.tex
\begin{tikzpicture} [every mark/.append style={mark size=1pt}]
\pgfplotsset{every tick label/.append style={font=\scriptsize}}
\definecolor{color0}{rgb}{0.12156862745098,0.466666666666667,0.705882352941177}
\definecolor{color1}{rgb}{1,0.498039215686275,0.0549019607843137}

\begin{axis}[
width=\fwidth,
height=\fheight,
at={(0\fwidth,0\fheight)},
scale only axis,
axis line style={white!15!black},
x grid style={white!80!black},
xlabel={Number of vehicles},
xmajorgrids,
xmajorticks=true,
xmin=4, xmax=21,
xtick style={color=white!15!black},
x label style={font=\footnotesize,at={(axis description cs:0.5,-0.13)},anchor=north},
y label style={font=\footnotesize,at={(axis description cs:-0.09,0.5)},anchor=south},
y grid style={white!80!black},
ylabel={Average latency [ms]},
ymajorgrids,
ymajorticks=true,
ymin=0.6, ymax=2,
ymode=log,
ytick style={color=white!15!black},
ytick={0.6,1,1.5},
yticklabels={0.6,1,1.5},
legend style={legend cell align=left, align=left, draw=white!15!black,
anchor=east, font=\scriptsize, align=left, at={(0.99,0.5)}},
legend columns=1,
]
\addplot [line width=1pt, color0, mark=diamond]
table {%
4 0.724023706666667
5 0.69138622
6 0.67345855
7 0.67968648
8 0.677044456666667
9 0.678535836666667
10 0.66578154
11 0.672170966666667
12 0.677343677083333
13 0.6611840725
14 0.677512647083333
15 0.677531556666667
16 0.679980166666667
17 0.679702
18 0.679100816666667
19 0.683737056666667
20 0.68723170125
21 0.711228523333333
};
\addlegendentry{UAV---BS ($L1$)};
\addplot [line width=1pt, color1, mark=*, densely dashed, mark options={solid}]
table {%
4 1.02094194262295
5 1.075554
6 1.29350756284153
7 1.52001036533958
8 1.52982475819672
9 1.53806621129326
10 1.54957120327869
11 1.55701031594635
12 1.5884823227459
13 1.61247813871375
14 1.637242470726
15 1.64546474262295
16 1.62822969057377
17 1.62543023336548
18 1.67275144626594
19 1.72290280759275
20 1.7758732295082
21 1.82658944574551
};
\addlegendentry{BS---Vehicles ($L2$)};

\end{axis}

\end{tikzpicture}